\documentclass{epl}

\newcommand{\MyRe}{{\cal R} e \,}

\newcommand{\lsim}{\raisebox{-1mm}{{\scriptsize $\,\stackrel{\textstyle <}{\textstyle \sim}\,$}}}
\newcommand{\gsim}{\raisebox{-1mm}{{\scriptsize $\,\stackrel{\textstyle >}{\textstyle \sim}\,$}}}

\title{Correlated tunneling into a superconductor in a multiprobe hybrid 
structure}

\shorttitle{Correlated tunneling into a superconductor}
\author{G. Falci\inst{1,2,3} \and D. Feinberg\inst{3} \and
F.W.J. Hekking\inst{4}}
\shortauthor{G. Falci \etal}
\institute{
  \inst{1} Dipartimento di Metodologie Fisiche e Chimiche (DMFCI),
Universit\`a di Catania -  viale A. Doria 6, 95125 Catania, Italy. \\
\inst{2}
Istituto Nazionale per la Fisica della Materia (INFM), Unit\`a di Catania.
\\
  \inst{3} Laboratoire d'Etudes des Propri\'et\'es Electroniques
des Solides, Centre National de la Recherche Scientifique, under
contract with Universit\'e Joseph Fourier, BP 166, 38042 Grenoble Cedex
9, Grenoble, France
\\
 \inst{4} Laboratoire de Physique et Mod\'elisation des Milieux
Condens\'es, Centre National de la Recherche Scientifique and
Universit\'e Joseph Fourier,
BP 166, 38042 Grenoble Cedex 9, Grenoble, France
}

\begin{document}

\maketitle

\begin{abstract}
We consider tunneling in a hybrid system consisting of a 
superconductor with two or more probe electrodes which can be either normal 
metals or polarized ferromagnets.
In particular we study transport at subgap voltages and temperatures. 
Besides Andreev pair tunneling at each contact, in multi-probe structures  
subgap transport involves additional channels, which are due to 
coherent propagation of two particles (electrons or holes), each originating 
from a different probe electrode. 
The relevant processes are electron cotunneling through the superconductor 
and conversion into a Cooper 
pair of two electrons stemming from different probes. 
These processes are non-local and
decay when the distance between the pair of involved contacts is larger than
the superconducting coherence length. The conductance
matrix of a three terminal hybrid structure is calculated. The multi-probe
processes enhance the conductance of each contact. If the contacts
are magnetically polarized  the contribution of the various conduction 
channels can be separately detected.
\end{abstract}

Superconductor-normal metal (SN) 
contacts at mesoscopic scale are of primary importance
in view of the interplay between coherence effects in the metal and
intrinsic coherence of the superconducting condensate which is probed by
Andreev reflection~\cite{Andreev,BTK}.
The situation becomes even more interesting with ferromagnetic metals (F).
The subgap Andreev conductance tends to be hindered by magnetic 
polarization~\cite{deJBeen,Soul,Upad} of the F electrode, since not every 
electron from the spin up band can find a spin down partner to be converted
in a Cooper pair. Recent experiments~\cite{kn:Giroud98,kn:Petrashov99} 
suggesting a proximity effect in SF junctions are still under debate.
More generally, multi-terminal NS structures offer the possibility of 
manipulating 
phase coherent transport~\cite{LambertRaim}.  
An example is the Andreev interferometer consisting of
a mesoscopic N sample, made of two arms connected to two S ``mirrors'' on 
one side and to a single reservoir on the other,    
where electron propagation in N is sensitive to the difference of the phase 
of the  two superconductors. In general the two SN contacts are separated by 
a distance larger than the superconducting coherence length 
$\xi = \hbar v_F/\pi\Delta$, $\Delta$ being the superconducting
energy gap. 
A dual configuration, consisting in a superconductor 
connected to two N probe electrodes separated by a 
distance {\it smaller} than $\xi$ (see Fig 1.a), with two
{\it independent} reservoirs,  
was considered in Ref.~\cite{Byers,us}.
It has been proposed that correlations between the N probes could be 
established across the superconductor, by a process where
two electrons {\em each originating from a different N electrode}, 
are converted in a 
Cooper pair. Recently, it has 
been shown that such correlations could be used to build entangled 
states of electrons \cite{Martin,Choi}. 
If the probe electrodes are ferromagnetic, the conductance 
due to these Crossed Andreev (CA) processes is sensitive to the relative 
magnetic polarization, 
being maximal for opposite polarization of the F probes~\cite{us}. 

In the present Letter, we investigate in detail the dependence of the 
conductance on the distance between the probes and on their spin 
polarization. We consider a three-terminal device $A/S/B$
(see Fig.1a) where $S$ is an s-wave superconductor
and the electrodes $A$ and $B$ can be either normal metals or ferromagnets.
We study the linear conductance matrix at subgap temperature and 
voltages (the subgap regime $T,eV_i \ll \Delta$ is assumed in what follows).
Then no single electron channel is left and all the relevant channels 
involve simultaneous (on a time $\sim \hbar/\Delta$) tunneling of two 
electrons or holes (Fig. 1b) :
(a) single-contact Andreev reflection 
($2A$ and $2B$), where two electrons, both originating from the same electrode,
$A$ or $B$, are converted in a Cooper pair giving rise to the currents 
$G_{2A}V_A$ and $G_{2B} V_B$;
(b) CA processes, where  two electrons,
each originating from a different $A/B$ electrode are converted into a
Cooper pair, with associated current $G_{CA} (V_A + V_B)$ from each $A/B$
probe to $S$;   
(c) cotunneling (EC), which is easily visualized in the tunneling 
limit~\cite{AverinNazarov} as processes in which, for instance, 
an electron from A tunnels to B via a virtual state in S;
EC processes yield a current from $A$ to $B$, 
$G_{EC}(V_B - V_A)$.
Then the subgap currents can be presented in a matrix form
\begin{equation}
\label{eq:current}
\left(
\begin{array}{c}
I_A\\
I_B \rule[12pt]{0pt}{1pt}
\end{array}
\right) \;=\;
\left(
\begin{array}{cc}
G_{2A} +  G_{CA} + G_{EC} & G_{CA} -  G_{EC}\\
 G_{CA} -   G_{EC} & G_{2B} +  G_{CA} + G_{EC}
\rule[12pt]{0pt}{1pt}
\end{array}
\right)
\;
\left(
\begin{array}{c}
V_A\\
V_B \rule[12pt]{0pt}{1pt}
\end{array}
\right)
\end{equation}
The multi-contact processes, CA and EC, increase the "diagonal" 
conductances $dI_A/dV_A, dI_B/dV_B$ and 
give rise to off-diagonal terms $dI_A/dV_B, dI_B/dV_A$, 
i.e. the current at probe $A$  ($B$) 
depends also on the voltage at probe $B$ ($A$). The generalization of 
Eq.(\ref{eq:current}) to more complicated structures (see Fig.1.c) is 
straightforward. 

For illustrative purposes we calculate the conductance matrix 
for contacts being tunnel junctions. The single-junction conductances  
$G_{2A}$ and $G_{2B}$ were calculated in  Ref.~\cite{HekkingNazarov}.
Here we find that EC and CA conductances depend on the relative position 
$\vec{R}$  of the contacts, vanishing exponentially for 
$|\vec{R}| \gg \xi$. For multichannel tunnel junctions they are 
found to depend on propagation in $S$ in a way which is sensitive to 
the geometry of the sample. This is due to interference effects 
between different channels, which also play a role in determining 
the single-junction NS conductances~\cite{HekkingNazarov}.
In  multichannel junctions with normal $A/B$ electrodes we find 
$G_{CA}=G_{EC}$, thus the off-diagonal terms in the conductance 
matrix Eq.(\ref{eq:current}) vanish.
This symmetry is broken if $A/B$ are polarized ferromagnets. Indeed  $G_{EC}$
is suppressed if  $A$ and $B$ have opposite polarization, because 
EC processes preserve the spin of the involved electron (in absence of 
any magnetic scattering whatsoever) and 
take advantage from parallel  $A/B$ polarization. 
On the other hand CA tunneling is suppressed for parallel $A/B$ polarization
since it is difficult to find two partner electrons with opposite spin to
pair up.  Then if magnetically polarized probes are used, off-diagonal 
conductances in Eq.(\ref{eq:current}) are non vanishing, due to the presence
of unbalanced CA and EC processes, for inter-contact distance 
$|\vec{R}| \lsim \xi$.

\section{Cotunneling and Crossed Andreev tunneling rates}
We describe contacts in the tunneling regime 
by standard tunneling Hamiltonians
\begin{eqnarray}
{\cal H}_{TA} \;=\;
\sum_{kp\sigma} \; T^A_{kp} \; c^{\dagger}_{k\sigma} d_{p\sigma} \,+\,
T^{(A)*}_{kp} \; d^{\dagger}_{p\sigma} c_{k\sigma}
\quad ; \quad
{\cal H}_{TB} \;=\;
\sum_{pq\sigma} \; T^{B}_{pq} \; d^{\dagger}_{p\sigma} c_{q\sigma} \,+\,
T^{(B)*}_{pq} \; c^{\dagger}_{q\sigma} d_{p\sigma}
\end{eqnarray}
where $T^A_{kp}$ and $T^{B}_{qp}$ are matrix elements
between single electron states $k \in A$,  $p \in S$ and $q \in B$.
Quasiparticle states in $S$ are defined by the operators $\gamma_{p
\sigma} =  u_{p \sigma} d^+_{p \sigma} - v_{p \sigma} d_{-p -\sigma}$.

\begin{figure}
\label{figure}
\onefigure{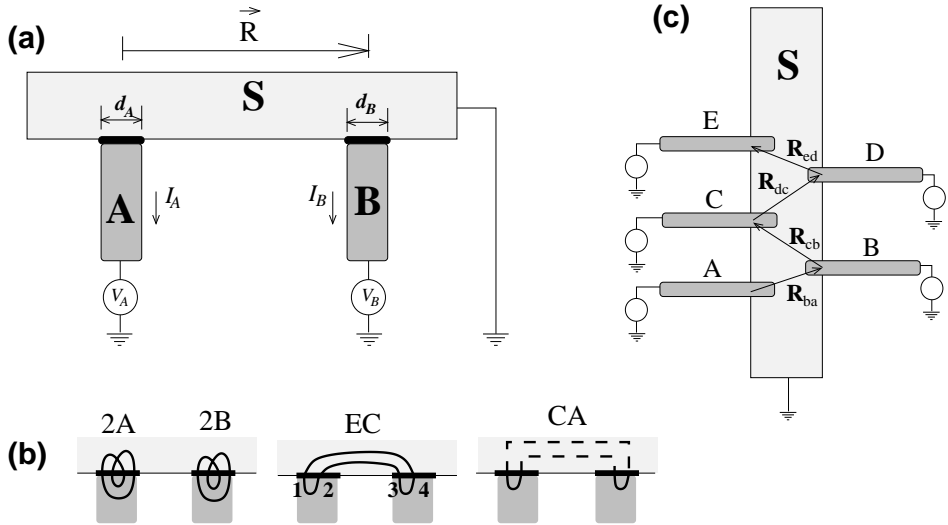}
\caption{(a) Schematics of the three-terminal A/S/B device;
(b) diagrammatic representation of the processes leading to subgap
conductance: single-contact two-particle tunneling ($2A$ and $2B$),
elastic cotunneling (EC), which probes the normal Green's function 
(full lines) of
$S$ and ``crossed Andreev'' (CA), which probes the anomalous propagator
(dashed line);
(c) a simple experimental design for measuring the excess EC and CA
currents: various electrodes ($A$,$B$,$C$,$D$,$E$) allow to probe
different distances $\vec{R}_{ij}$.}
\end{figure}

We now classify processes which appear in perturbation theory in
${\cal H}_{Ti}$.
Single electron processes are absent in the subgap regime.
In lowest nonvanishing order we have to consider only ``elastic''
processes, where the quasiparticle created in the intermediate state is
destroyed~\cite{AverinNazarov}.
``Inelastic'' processes leave an excitation of energy $\,\gsim \Delta\,$  
in the superconductor, so under subgap conditions they can be neglected. 
The relevant processes, represented in
Fig.1b, are 
single-contact Andreev  reflection in the tunneling limit 
($2A$ and $2B$)~\cite{HekkingNazarov}, CA tunneling, and 
``elastic'' cotunneling~\cite{AverinNazarov}, where {\em the same} 
electron tunnels at once from $A$ to $B$ ($B$ to $A$). 
One notices that
EC and CA are non-local probes of electron propagation in
the superconductor\cite{Byers}. This is apparent if one considers the
gedanken case
of single channel tunnel junctions, where the size of the contacts is 
$d_A,d_B\sim \lambda_F$, the Fermi
length.
EC probes the "normal" propagator in the superconductor
whereas CA processes probe the {\it anomalous} propagator
(see Fig.1b) and  both processes are
exponentially suppressed for $R \gg \xi$.

We now turn to the calculation of the spin dependent tunneling rates, using
Fermi's golden rule. 
We consider electrodes $A$ and $B$ which can be magnetically
polarized (parallel or antiparallel) along the same axis, resulting in
different spectral properties. We neglect here the
influence of ferromagnetic electrodes on the
superconductor's spectrum. This a reasonable assumption for small tunneling 
contacts on a massive superconductor. In this case, the proximity effect 
should not be too serious, e. g. superconductivity is not destroyed in 
the contact vicinity. This is in contrast with 
Ref. \cite{Takahashi} where a sandwich geometry is considered, and where 
conduction above the gap is studied. 

By proceeding along the lines of Ref. \cite{HekkingNazarov}
we write the  EC rate $A\sigma \to B\sigma$ as
\begin{eqnarray}
\label{eq:rate-EC}
\Gamma_{A\to B}^{\sigma} &=& {2 \pi \over \hbar} \;
\int \!\! d\varepsilon d\varepsilon^{\prime} d\zeta d\zeta^{\prime} \;
\delta(\varepsilon- \varepsilon^{\prime}) \;
f(\varepsilon-e V_A) \; [1- f(\varepsilon^{\prime}-e V_B)]
\nonumber
\\&& \hskip40mm
{\mathrm{F}}_{EC}(\zeta, \varepsilon)
\; {\mathrm{F}}_{EC}(\zeta^{\prime}, \varepsilon^{\prime})
\;\;
\Xi_{EC}^{\sigma}(\varepsilon-e V_A, \varepsilon^{\prime}-e
V_B,\zeta,\zeta^{\prime})
\hskip5mm \end{eqnarray}
where $f(\varepsilon)$ is the Fermi function,
${\mathrm{F}}_{EC}(\zeta, \varepsilon) \;=\;
(\zeta + \varepsilon)/ (\zeta^2 + \Delta^2 - \varepsilon^2)$
contains information on energies of virtual quasiparticles in $S$.
The rate for CA processes,  $(A\sigma, B-\sigma)\to S$, is given by
a similar expression
\begin{eqnarray}
\label{eq:rate-CA}
\Gamma_{AB \to S}^{\sigma} &=& {2 \pi \over \hbar} \;
\int \!\! d\varepsilon d\varepsilon^{\prime} d\zeta d\zeta^{\prime} \;
\delta(\varepsilon + \varepsilon^{\prime}) \;
f(\varepsilon-e V_A) \, {f}(\varepsilon^{\prime}-e V_B)
\nonumber
\\&& \hskip40mm
{\mathrm{F}}_{CA}(\zeta, \varepsilon)
\; {\mathrm{F}}_{CA}(\zeta^{\prime}, \varepsilon^{\prime})
\;\;
\Xi_{CA}^{\sigma}(\varepsilon-e V_A, \varepsilon^{\prime}-e
V_B,\zeta,\zeta^{\prime})
\hskip5mm
\end{eqnarray}
where $
{\mathrm{F}}_{CA}(\zeta, \varepsilon) \;=\;
\Delta/ (\zeta^2 + \Delta^2 - \varepsilon^2)$. Information about propagation
in the specific geometry are contained in the functions
$\Xi(\varepsilon, \varepsilon^{\prime},\zeta,\zeta^{\prime})$. Here we
give the explicit expression for planar uniform tunnel junctions
and local tunneling, $T(\vec{r},\vec{r^{\prime}}) \,=\, T \,\delta(z) \,
\delta (\vec{r}-\vec{r^{\prime}})$. We moreover consider ballistic
propagation of plane wave
states (generalization to diffusive conductors can be made following
Ref.\cite{HekkingNazarov}). In this case the functions $\Xi_{EC}$
and $\Xi_{CA}$ can be expressed as
\begin{eqnarray}
\label{eq:Xi-funct}
\Xi_{EC}^{\sigma}(\varepsilon, \varepsilon^{\prime},\zeta,\zeta^{\prime}) &=&
|T_A T_B|^2 \int_{A} \hskip-1mm d\vec{r}_1 d\vec{r}_2 \;
\int_{B} \hskip-1mm d\vec{r}_3 d\vec{r}_4  \; \;
\rm{J}_{A}^{\sigma}(12,\varepsilon)\,
\,\rm{J}_S^{\sigma}(31,\zeta)\,
\rm{J}_S^{\sigma}(24,\zeta^{\prime})\,
\rm{J}_{B}^{\sigma}(43, \varepsilon^{\prime})\\
\nonumber
\Xi_{CA}^{\sigma}(\varepsilon, \varepsilon^{\prime},\zeta,\zeta^{\prime}) &=&
|T_A T_B|^2 \int_{A} \hskip-1mm d\vec{r}_1 d\vec{r}_2 \;
\int_{B} \hskip-1mm d\vec{r}_3 d\vec{r}_4  \; \;
\rm{J}_{A}^{\sigma}(12,\varepsilon)\,
\,\rm{K}_S^{\sigma}(31,\zeta)\,
\rm{K}_S^{*\sigma}(24,\zeta^{\prime})\,
\rm{J}_{B}^{-\sigma}(43, \varepsilon^{\prime})
\end{eqnarray}
where the spectral functions are defined as, e.g.
$\rm{J}_A^{\sigma}(12, \omega) \equiv
\rm{J}_A^{\sigma}(\vec{r}_1,\vec{r}_2, \omega)
\,=\, \sum_k \delta(\omega-\varepsilon_{\mathbf{k}{\sigma}}) \,
\psi_{k\sigma}(\vec{r}_1) \psi_{k\sigma}^*(\vec{r}_2)$, 
$\rm{K}_S^{\sigma}(31, \omega) \equiv
\rm{K}_A^{\sigma}(\vec{r}_3,\vec{r}_1, \omega)
\,=\, \sum_k \delta(\omega-\varepsilon_{\mathbf{k}{\sigma}}) \,
\psi_{-k -\sigma}(\vec{r}_3) \psi_{k \sigma}(\vec{r}_1)$.
The space
integrals in (\ref{eq:Xi-funct}) run on the contact surfaces.
The diagrammatic representation is
given in Fig.~1.b.

At low temperature and voltages the main contribution to the rates
Eqs.(\ref{eq:rate-EC},\ref{eq:rate-CA}) is due to electrons 
close to the Fermi level in the $A/B$ probes, 
$\varepsilon = \pm \varepsilon^\prime \approx 0$, and the  
leading dependence of the rates on the voltages comes from the 
Fermi functions. 
One can let 
$\Xi(\varepsilon, \varepsilon^{\prime},\zeta,\zeta^{\prime})\approx
\Xi(0, 0,\zeta,\zeta^{\prime})$ and 
perform the $\varepsilon$ and $\varepsilon^\prime$ integrations in 
Eqs.(\ref{eq:rate-EC},\ref{eq:rate-CA}). At $T=0$ this gives
$I_{EC}^{\sigma} = e \Gamma_{A\to B}^{\sigma} = G_{EC}^{\sigma} (V_B - V_A)$
and
$I_{CA}^{\sigma} = 2  e \Gamma_{AB \to S}^{\sigma} =
2 G_{CA}^{\sigma} (V_A + V_B)$, 
which define the spin-dependent conductances.

It is instructive to consider first single channel junctions
($d_{A}, d_{B} \sim \lambda_F$). The conductances are 
calculated by putting $\vec{r}_1=\vec{r}_2$,
$\vec{r}_3=\vec{r}_4$, $\vec{r}_1 - \vec{r}_3 = \vec{R}$
in Eq.(\ref{eq:Xi-funct}). By performing the $\zeta$ and $\zeta^{\prime}$
integrations the result is obtained
\begin{eqnarray}
\label{point-contacts}
\left(
\begin{array}{c}
G_{EC}^{\sigma}  \\
G_{CA}^{\sigma} \rule[5mm]{0mm}{0.0pt}
\end{array}
\right)
\;\approx\;
{2 \pi^3 e^2 \over \hbar} \; |T_A d_A|^2 \, |T_B d_B|^2   \;
N_{S}^2(0)  N_A^{\sigma}(0)  \;
{{\mathrm{e}}^{- 2R/\pi\xi}
	\over  (k_{S} R)^2}\;
\;
\left(
\begin{array}{c}
N_{B}^{\sigma}(0) \; \cos^2(k_S R)\\
N_{B}^{-\sigma}(0) \;\sin^2(k_S R) \rule[5mm]{0mm}{0.0pt}
\end{array}
\right)
\end{eqnarray}
where $k_S$ and $N_S(0)$ are the Fermi wavevector and the normal state 
density of states of the superconductor. 
Propagation in the superconductor is characterized by factors which
depend on $|\vec{R}|$, being periodic with period $\pi/k_S$ and being 
suppressed for $|\vec{R}| \gg \xi$.
The magnetic polarization of $A/B$ electrodes enters only via the 
spin-dependent density of states
$N_{A,B}^{\sigma}(0)$, assuming that $N_{A,B}^{\sigma}(0) = N_{A,B}^0(\pm 
\mu h)$ where $h$ is the exchange field, $N_{A,B}^0$ is the density of states in 
absence of magnetism and $\mp$ sign stands for spin $\sigma$ 
(anti)parallel to the magnetization. If the density of states for minority spin can be
neglected, then for parallel
(antiparallel) polarized contacts only EC (CA) processes will yield a
conductance.

For the more realistic case of multichannel tunnel junctions,
interference between different channels has to be considered.
This has been studied in Ref.\cite{HekkingNazarov} for
single contact processes ($A1$ and $A2$) in a $NS$ tunnel junction, 
which probe propagation in the {\em normal} electrode
(it is in general diffusive and depends on the geometry).
As for EC and CA, we notice that in clean superconductors
the single particle propagator is rapidly oscillating
($\sim k_S$), so it would average out on a length $\xi \ll 1/k_S$
in  multichannel junctions.
This is not the case for two-particle propagators involved in 
EC and CA processes, as we show explicitly below by considering 
a specific geometry.

At this stage one could assume that the overall conductance 
is given by independent single-channel contributions, 
start from Eq.({\ref{point-contacts}) and argue that the 
factors $\cos^2(k_S R)$ and $\sin^2(k_S R)$ are averaged over distances 
$\sim d_A,d_B$. The resulting expressions for the 
EC and CA conductances would be nearly identical yielding, 
in the special case of normal $A/B$ electrodes, 
$G_{EC}=G_{CA}$.
This conclusion turns out to be correct, even if the actual expression 
for the conductances involves interference between different channels.
To show that we start the calculation from Eq.(\ref{eq:Xi-funct}), 
accounting also for different spin-dependent Fermi
wavevectors $k_{A}^{\sigma}$ ($k_{B}^{\sigma}$) of the $A$ ($B$)
electrode and $k_S$.
We perform the coordinate integrations, and 
at this stage it becomes apparent that the terms leading to the 
special dependence $\cos^2(k_S R)$ and $\sin^2(k_S R)$ drop out. 
Because of interference between different channels, the result depends on 
the geometry and on the mismatch between the
spin-dependent Fermi wavevectors. The simplest case is a geometry with the 
two junctions belonging to the same plane (see Fig.1a), where the results 
depend on the distance $R > d_A,d_B$
\begin{eqnarray}
\label{EC-CA-conductance}
\left(
\begin{array}{c}
G_{EC}^{\sigma}  \\
G_{CA}^{\sigma} \rule[5mm]{0mm}{0.0pt}
\end{array}
\right)
\;\approx\;
{h \over 8 e^2} \; {\cal F}_A^{\sigma}
\; G_A^{\sigma}
\left(
\begin{array}{c}
{\cal F}_B^{\sigma} \, G_{B}^{\sigma}  \\
{\cal F}_B^{-\sigma} \, G_{B}^{-\sigma} \rule[5mm]{0mm}{0.0pt}
\end{array}
\right)
\;
{{\mathrm{e}}^{- 2R/\pi\xi}
	\over  (k_{S} R)^2}\;
\end{eqnarray}
Here the factors ${\cal F}_{A,B}^{\sigma}$ contain information on the
geometry and $G_{A,B}^{\sigma}$ are the spin-dependent
one-electron conductances for each junction (S being in the normal state),
for instance
\begin{eqnarray}
\label{eq:planar-NN-conductance}
G_{A}^{\sigma} \;\approx\;
{4 \pi e^2 \over \hbar}\;  N_{A}^{\sigma}(0)  N_S(0)\;
\;{\mid T_A \mid^2 \, {\cal S} \over  k_{A}^{\sigma}  k_{S} }\;\;
{\cal F}(\kappa_{A}^2,\sqrt{k_{S}k_{A}^{\sigma}}d_{A})
\end{eqnarray}
where
$\kappa_{A,B} = (k_{A,B}^{\sigma}/k_S)^{1/2}\,$ and ${\cal S}$ is the
area of the junction.
Here the single junction geometry factor is given by
${\cal F}(\kappa^2,y)
= 2 \pi  \int_0^{y} (dx/x) \, \sin (\kappa x)\, \sin (x/\kappa)\,$
and 
the two-junction factors in Eq.(\ref{EC-CA-conductance}) are, e.g.,
${\cal F}_A \,=\, [{\cal F}(\kappa_{A}^2,\sqrt{k_{S}k_{A}}d_{A})]^{-1} \,
\MyRe \hskip-1mm \int_0^{2\pi} \hskip-1mm d \theta \,
[ 1 - {\mathrm{e}}^{i k_S d_A (\kappa^2_A+\cos \theta)}]/
(\kappa^2_A+\cos \theta)\,$ for
the geometry we consider. It is important to point out only
some general property. For  $\kappa = 1$ the factor
${\cal F}(1,y) \propto \ln y $ depends weakly\cite{kn:Hekking}
on the reduced size $y$; the factors ${\cal F}_{A,B}$ are
even more weakly dependent on  $y$ and substantially of order one.
A slight asymmetry  $\kappa \neq 1$ makes all the factors $\cal F$ 
independent on the size of the junctions,  if $d_A,d_B$
are large enough, $y |\kappa - \kappa^{-1}| \gg 1$. Still
${\cal F}_{A,B}$ are of order one,
so EC and CA processes determine an appreciable conductance.

\section{Discussion}
We can now discuss the full conductance matrix in equation (\ref{eq:current})
by defining the total EC and CA conductivities $G_{EC} = G_{EC}^{\sigma} +
G_{EC}^{-\sigma}$ and $G_{CA} = G_{CA}^{\sigma} + G_{CA}^{-\sigma}$.
The EC and CA conductances appear both in the diagonal
and in the off-diagonal conductance matrix elements in Eq.(\ref{eq:current}).
Let us first consider the case of non magnetic contacts, where we can
drop the spin dependence.
For multichannel contacts Eq.(\ref{EC-CA-conductance}) shows that
$G_{EC} = G_{CA}\,$, so the off diagonal conductances {\it vanish}.
Coherent tunneling processes involving
two distant contacts enter only the diagonal terms, and 
provide as an extra contribution with respect to the standard 
Andreev conductances $G_{2A}$ and $G_{2B}$. The extra current
depends on the distance $R$ between the two
contacts. A simple setup where the $R$ dependence of
the extra current can be studied is shown in Fig.1c
(alternatively one may use a STM tip as a mobile contact).
Another signature of CA and EC processes can be found if one considers
contacts of very different transparency, say $|T_A| \gg |T_B|$. 
In this case $G_{2A} \;\propto \; |T_A|^4$ dominates
$G_{EC} + G_{CA} \propto |T_A|^2 |T_B|^2$, which
itself is much larger than $G_{2B} \;\approx \; |T_B|^4$. 
Thus the conductance at the less transparent probe 
(lower right diagonal element in Eq.(\ref{EC-CA-conductance})) is 
given by  $G_{EC} + G_{CA}$, so it is essentially due to 
two-contact processes: if we bias contact $B$ a current will flow  
because of correlations of superconductive nature with contact $A$.   
If $R < \xi$, the crossed conductances are still affected by the factor 
$(k_S R)^{-2}$ which can be very small \cite{Choi}. This problem can be 
partially overcome for instance by choosing one contact A to be of size 
$d_A>>\xi$ or 
even a semi-infinite interface.

Let us now consider the case of spin polarized probes.
As it is apparent from Eq.(\ref{EC-CA-conductance}) $G_{EC} \neq G_{CA}$
so the off diagonal elements in the conductance matrix, Eq.(\ref{eq:current}), 
are finite and the current in one contact can be manipulated by the voltage
bias of the other contact. The sign of this effect depends on the mutual 
polarization of the electrodes. 
This generalizes the result of Ref.\cite{us}. More in detail,
spin polarization enters in
two ways in the result (\ref{point-contacts},\ref{EC-CA-conductance}):
first, in the spin-dependent densities of states;
second, in the shift of the Fermi momenta $k_{A,B}^{\sigma}$, which
modifies the factors $\cal F$ in Eqs.(\ref{EC-CA-conductance}).
To fix the ideas, let us for simplicity neglect the latter and
concentrate on the effect of the density of states.
Defining the contact polarizations $P_{A,B} =
\frac{N_{A,B}^{\sigma} - N_{A,B}^{-\sigma}}{N_{A,B}^{\sigma} +
N_{A,B}^{-\sigma}}$, one has simply that $G_{EC}$ is proportional to $(1 +
P_A P_B)$ and $G_{CA}$ to $(1 - P_A P_B)$. Therefore the off-diagonal
conductance is roughly proportional to $(-P_A P_B)$. This shows a striking
consequence of the competition between cotunneling and crossed Andreev
processes, via their spin-dependence : non only the
amplitude, but also the sign of the conductances
can be controlled by spin polarizations. In the extreme case of parallel
complete polarizations the only possible process is cotunneling,
with $I_A =  -I_B$, while for antiparallel polarization crossed Andreev 
tunneling prevails, with $I_A = I_B$.

So far we have discussed the zero-temperature case. Direct generalization
to finite temperature leads formally to a divergence of the
tunneling rate. It is due to
the finite, though very small ($\propto e^{-\Delta/T}$), probability of
exciting an electron from $A(B)$ to a quasiparticle state in S, and to the
divergence of the quasiparticle density of states in S.
The divergence in the rates disappears if the latter is
rounded off at $\Delta$. The EC and CA conductances acquire an 
additional contribution 
$\propto e^{-\Delta/T}\, \ln(\Delta /\Gamma)$ where
$\Gamma$ is a scale related to the mechanism of broadening of the
quasiparticle levels in the superconductor.

In the present Letter we have demonstrated the non-local character of
cotunneling and Andreev reflections on a superconductor and we have also 
studied the role of magnetic polarization. 
We have discussed possible schemes to detect these effects in devices with 
three or more terminal. Devices with high transparency 
contacts~\cite{us} are also promising for  experiments.
Further theoretical analysis would require the self-consistent analysis of 
the mutual effects of superconductivity, diffusive propagation and 
ferromagnetism in the hybrid system. For both high and low transparency 
contacts propagation in the specific geometry has to be taken into account.

\acknowledgments
One of the authors (D. F.) is grateful to Prof. G. Deutscher for
stimulating discussions. G.F. acknowledges discussions with 
R. Fazio, support and hospitality by
LEPES-CNRS Grenoble (France), and support from  MURST-Italy
(Cofinanziamento SCQBD), INFM (PAIS-ELMAMES) and
EU (grant TMR-FMRX-CT-97-0143).

\end{document}